\documentclass[aps,prl,twocolumn,superscriptaddress,showpacs,draft]{revtex4}
\usepackage{graphicx}
\input{epsf}

\begin{document}

\title{Anderson transition in three and four effective dimensions \\
for the frequency modulated kicked rotator}

\author{D.L.Shepelyansky}
%\homepage[]{http://www.quantware.ups-tlse.fr}
\affiliation{\mbox{Laboratoire de Physique Th\'eorique du CNRS (IRSAMC), 
Universit\'e de Toulouse, UPS, F-31062 Toulouse, France}}
%\affiliation{\mbox{LPT (IRSAMC), CNRS, F-31062 Toulouse, France}}

%\date{\today}
\date{ February 21, 2011}

%\pacs{PACS numbers: 05.45.-a, 05.45.Ac, 05.45.Jn}
%\PACS{
%{05.45.-a}{Nonlinear dynamics and chaos}
%\and
%{05.45.Ac}{Low-dimensional chaos}
%{05.45.Jn}{High-dimensional chaos}
%}
%03.75.-b 	Matter waves 
%05.45.Mt 	Quantum chaos; semiclassical methods 
%72.15.Rn 	Localization effects (Anderson or weak localization) 

\pacs{05.45.Mt, 72.15.Rn, 03.75.-b}
\begin{abstract}
The critical exponents for the Anderson transition in three and four
effective dimensions are discussed on 
the basis of previous data obtained for the frequency modulated kicked rotator.
Without appeal to a scaling function they are shown to be in a satisfactory 
agreement with the theoretical relation known for them.
\end{abstract}

\maketitle

Recent experiments of the Garreau group with kicked cold atoms
\cite{garreau1,garreau2,garreau3}
renewed interest to the frequency modulated kicked rotator (FMKR), 
which had been introduced and studied some time ago
\cite{dls1983,dls1989,bs1996,bs1997}. 
The evolution of the quantum system is described
by the unitary propagation operator
\begin{equation}
\begin{array}{c}
{\hat S_1} = \exp ( -i  H_{0}({\hat n})) \exp (-i V(\theta,t))
\end{array}
\label{eq1}
\end{equation}
with quasiperiodic frequency modulation of kick potential
$V(\theta,t)=V(\theta, \theta_1, \theta_2)$ with
$\theta_{1,2} = \omega_{1,2} t$ at $d=3$
and $V(\theta,t)=V(\theta, \theta_1, \theta_2,\theta_3)$
with $\theta_{1,2,3} = \omega_{1,2,3} t$ at $d=4$.
Here, the notations following those of \cite{bs1996}.

The model with arctangent kick rotator (AKR) potential
was considered in \cite{dls1989,bs1997},
while the frequency modulated kicked rotator (FMKR)
corresponding to the quantum Chirikov standard map
with modulated kick amplitude had been analyzed in
\cite{dls1983,bs1996}. 

\begin{figure}
\centerline{\epsfxsize=7.0cm\epsffile{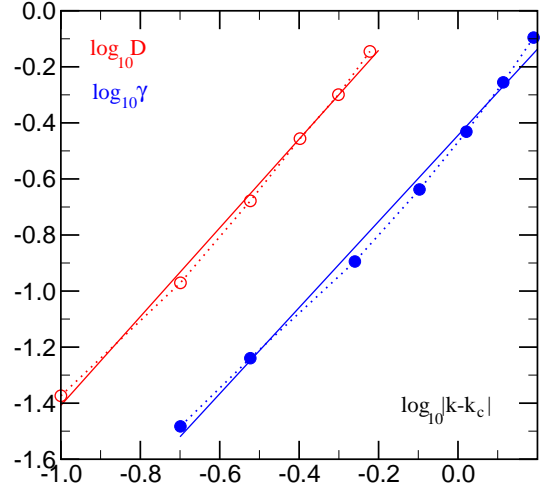}}
\vglue -0.2cm
\caption{(Color online) Power law dependence for the
inverse localization length $\gamma=1/l_1$ and diffusion rate $D$
obtained at asymptotically large times $t$ 
at FMKR with $\epsilon=0.75$
(in a vicinity of critical point $t \geq 10^6$).
Data are taken from Fig.1 of \cite{bs1996}
and are plotted for the fixed value of critical parameter
$k_c=1.8$ for dimension $d=3$ with full circles
for $\gamma$ and open circles for $D$
(dotted lines are drown to adapt an eye).
The straight lines show the power law fits
with fixed $k_c$ and
$\nu =1.537 \pm 0.0539$, 
$\log_{10} \gamma_0 = -0.444 \pm  0.0192$
and $s=1.583 \pm 0.0511$, $\log_{10} D_0 = 0.175 \pm  0.0301$.
} 
\label{fig1}
\end{figure}

In view of this interest I reconsider in more detail
the results presented in \cite{bs1996}
for FMKR in effective dimensions $d=3, 4$
(data of Figs.1,9 in \cite{bs1996} respectively).
In the case $d=3$ we have 
$V(\theta,t)=k \cos\theta (1+\epsilon \cos \omega_1t \cos \omega_2 t)$
with fixed $\epsilon=0.75$, irrational frequencies $\omega_{1,2}$
and random but fixed in time rotational phases $H_0(n)$
(see \cite{bs1996} for detailed notations).
The similar choice of $V(\theta,t)=
k \cos\theta (1+\epsilon \cos \omega_1t \cos \omega_2 t \cos \omega_3 t) $,
$\epsilon=0.9$ is done for the case of $d=4$ (data of Fig.9 in \cite{bs1996}).
Let me note that in the experiment \cite{garreau1,garreau2,garreau3}
the rotational phases correspond to a free propagation
with $H_0(n)=T n^2/2$ with the classical chaos
parameter $K=k T$. However, even for one modulation frequency
the chaos border for destruction of
two-frequency invariant torus is very low
so that the quantum chaotic dynamics 
mimics rather well random quantum phases $H_0(n)$.
Indeed, according to Fig.5 in \cite{artuso}
the invariant torus with two 
spiral mean frequencies is destroyed
at $K \approx 0.3$ for $\epsilon \approx 0.75$ that is significantly 
smaller than the experimental values with
$K \approx 6$, $T \approx 2.89$, $k \approx 2$. On the basis of these
arguments it is natural
to expect that the FMKR with random phases
has transition approximately 
at the same parameters as for the FMKR model 
with quadratic rotational phases (e.g. with $T=2$).
This was confirmed by the first numerical 
simulations for the case of quadratic rotational phases
performed in the proposal of Garreau
experiment in 2005 \cite{dls2005}
(done for effective $d=3$ at $\epsilon=0.75$, $T=2$
with estimated critical $k_c \approx 1.8$).
Hence, the both models are rather similar.
The transition border found in numerical simulations
\cite{garreau1,garreau2,garreau3}
is also in agreement with this statement.
Indeed, according to the data of Fig.1 in \cite{garreau1}
one finds $k=K/T=K/\hbar=1.88 \pm 0.035$
for $\epsilon=0.75$ and $T=\hbar=2.89$
that  agrees with the critical 
value $k_c=1.8$ given in \cite{bs1996}
for the case with random rotation phases.

The data of Figs.1,9 of \cite{bs1996} clearly show the 
presence of Anderson transition at $k_c \approx 1.8$ ($d=3$)
and $k_c \approx 1.15$ ($d=4$).
These data give the inverse localization length
$\gamma=1/l_1$ extracted from a steady-state 
distribution at asymptotically large times $t > 1/\gamma^d$
in the localized phase $k<k_c$. In the same way 
in the metallic phase they give the diffusion rate $D$
computed on asymptotically large times $t_a > 1/D^d$
from a gaussian distribution over sites.
The results are averaged over 100 disorder realizations for 
$k<k_c$ and 10 realizations for $k>k_c$ with $t \sim 10^6$.
In contrast to the approach used
in \cite{garreau1,garreau2,garreau3}, 
the extraction of such asymptotic values used in  \cite{bs1996}
does not rely on the existence of scaling function.
\begin{figure}
\centerline{\epsfxsize=7.0cm\epsffile{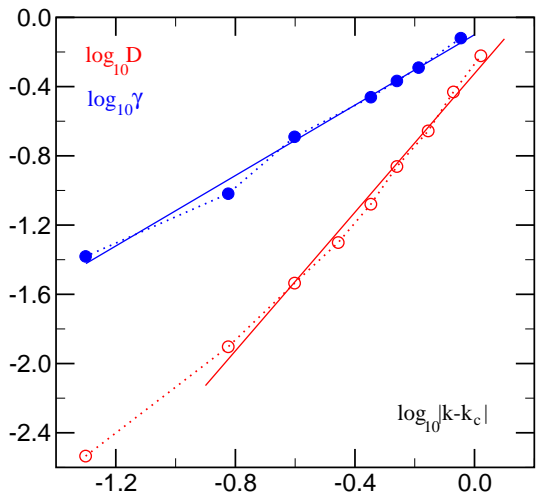}}
\vglue -0.2cm
\caption{(Color online) Power law dependence for the
inverse localization length $\gamma=1/l_1$ and diffusion rate $D$
obtained at asymptotically large times $t$ 
at FMKR with $\epsilon=0.9$
(in a vicinity of critical point $t \geq 10^6$).
Data are taken from Fig.9 of \cite{bs1996}
and are plotted for the fixed value of critical parameter
$k_c=1.15$ for dimension $d=4$ with full circles
for $\gamma$ and open circles for $D$
(dotted lines are drown to adapt an eye).
The straight lines show the power law fits
with fixed $k_c$ and
$\nu =1.017 \pm 0.041$, 
$\log_{10} \gamma_0 = -0.100 \pm  0.0266$
and $s=2.003 \pm 0.0740$,  $\log_{10} D_0 = -0.324 \pm  0.0317$
(the lowest value of $D$ is not taken into account in the fit
since the evolution time was smaller than $1/D^4$).
} 
\label{fig2}
\end{figure}

The data of \cite{bs1996} can be used for extraction of the
scaling exponents $\nu, s$ in the localized
$\gamma = \gamma_0 |k_c-k|^\nu$
and metallic $D = D_0 |k_c-k|^s$ phases.
For $d=3$ the fit of data for $\gamma$
with fixed critical point $k_c=1.8$
is shown in Fig.~\ref{fig1}. In a similar way the fit
for $d=4$ is shown in Fig.~\ref{fig2}.
The obtained values of $\nu$ and $s$ are in a satisfactory 
agreement with the scaling relation 
$s=(d-2)\nu$ (see. e.g. \cite{mirlin}) both for $d=3,4$.
For $d=3$ the values of $s$, $\nu$ are compatible
with those obtained in \cite{garreau1,garreau2,garreau3}.
According to \cite{delande}
D.Delande obtains for $d=4$
the critical exponents compatible with the
scaling relation and values similar to those  
of Fig.~\ref{fig2}.

Even if formally the statistical errors are
relatively small the actual values of $\nu$ and $s$
remain rather sensitive to variation of $k_c$.
Thus for $d=3$ a variation of $k_c$ by $\pm 0.05$
gives variation of $s$ by $+0.35, -0.40$ 
and of $\nu$ by $-0.17, +0.16$.
In a similar way for $d=4$
a variation of $k_c$ by $\pm 2\%$
gives variation of $s$ by $\mp 6\%$ 
and $\nu$ by $\pm 12\%$.
At the same time the statistical accuracy remains 
approximately on the same level.
The fit of data by 3-parameter power law
gives: $D_0=1.626 \pm 0.086$,
$k_c=1.728 \pm 0.039$, $s=2.074 \pm 0.177$ 
and $\gamma_0=0.135 \pm 0.0269$, 
$k_c=2.220 \pm 0.0783$, $\nu=2.626 \pm 0.145$
for $d=3$. Respectively, such a fit
for $d=4$ data gives
$D_0=0.397 \pm 0.090$,
$k_c=1.033 \pm 0.082$,
$s=2.697 \pm 0.263$
and $\gamma_0=0.821 \pm 0.028$,
$k_c=1.180 \pm 0.033$,
$\nu=1.176 \pm 0.091$.
These results show that the exact computation of the
critical exponents $s$ and $\nu$
remains very hard task even if the FMKR model
is much more efficient comparing to
the transfer matrix techniques and other numerical methods.
It is possible that the scaling methods
used in \cite{garreau1,garreau2,garreau3}
correspond to a better accuracy of the exponents.
However, this approach  uses extrapolation methods combined
with the scaling function which are not used in the 
data presented here.

A separate note should be done for the 
AKR model. The results presented in
\cite{bs1997} show very strong
deviation from the scaling relation between $s$ and $\nu$
for $4 \leq d \leq 11$. It is possible that at large $d$
the critical kick parameter becomes relatively small $k_c \sim 1/d$ and thus
this model becomes too close to a model with practically decoupled
quasi-periodic driving in time that can make it rather specific  or 
can require to study very small vicinity of $k_c$
for extraction of correct critical exponents.

\end{document}